\documentclass{article}

\usepackage{PRIMEarxiv}

\usepackage[utf8]{inputenc} 
\usepackage[T1]{fontenc}    
\usepackage{hyperref}       
\usepackage{url}            
\usepackage{booktabs}       
\usepackage{amsfonts}       
\usepackage{nicefrac}       
\usepackage{microtype}      
\usepackage{lipsum}
\usepackage{fancyhdr}       
\usepackage{graphicx}       
\usepackage{float}          
\usepackage{amsmath}        
\graphicspath{{media/}}     

\title{Stochastic approaches: modeling the probability of encounters between H$_{2}$-molecules and metallic atomic clusters in a cubic box}

\author{
  Maximiliano L. Riddick, Leandro Andrini \\
  Instituto de Investigaciones Fisicoquimicas Te\'oricas y Aplicadas \\
  Departamento de Qu\'imica, Fac. de Ciencias Exactas (INIFTA/ UNLP-CONICET) \\
  Departamento de Matem\'atica, Fac. de Ciencias Exactas, UNLP \\
  La Plata, Argentina\\
  \texttt{mriddick@mate.unlp.edu.ar} \\
  \AND
  Enrique E. \'Alvarez \\
   Instituto de C\'alculo, Fac. Ciencias Exactas y Naturales, Ciudad Universitaria, Pabell\'on II, UBA (CABA) \\
   Departamento de Fisicomatem\'atica, Fac. de Ingenier\'ia, UNLP \\
  Ciudad Autónoma de Buenos Aires and La Plata, Argentina \\
  \And
  F\'elix G. Requejo \\
  Instituto de Investigaciones Fisicoquimicas Te\'oricas y Aplicadas \\
  Departamento de Qu\'imica, Fac. de Ciencias Exactas (INIFTA/ UNLP-CONICET) \\
  Departamento de F\'isica, Fac. de Ciencias Exactas, UNLP \\
  La Plata, Argentina\\
}

\begin{document}
\maketitle

\begin{abstract}
In recent years the advance of chemical synthesis has made it possible to obtain \textquotedblleft naked\textquotedblright clusters of different transition metals. It is well known that cluster experiments allow studying the fundamental reactive behavior of catalytic materials in an environment that avoids the complications present in extended solid-phase research. In physicochemical terms, the question that arises is the chemical reduction of metallic clusters could be affected by the presence of H$_{_{2}}$ molecules, that is, by the probability of encounter that these small metal atomic agglomerates can have with these reducing species. Therefore, we consider the stochastic movement of $N$ molecules of hydrogen in a cubic box containing $M$ metallic atomic clusters in a confined region of the box. We use a Wiener process to simulate the stochastic process, with $\sigma$ given by the Maxwell-Boltzmann relationships, which enabled us to obtain an analytical expression for the probability density function. This expression is an exact expression, obtained under an original proposal outlined in this work, \textit{i.e.} obtained from considerations of \textit{mathematical rebounds}. On this basis, we obtained the probability of encounter for three different volumes, $0.1^{^{3}}$, $ 0.2^{^{3}} $ and $ 0.4^{^{3}} $ m$^{^{3}}$, at three different temperatures in each case, $293$, $373$ and $473$ K, for $ 10^{^{1}}\leq N \leq10^{^{10}}$, comparing the results with those obtained considering the distribution of the position as a Truncated Normal Distribution. Finally, we observe that the probability is significantly affected by the number $N$ of molecules and by the size of the box, not by the temperature.
\end{abstract}

\keywords{Wiener Process \and Probability of encounters \and Molecular Collisions \and Atomic-Clusters \and Mathematical Rebounds}

\section{Introduction}
In the last two decades there has been an important development in clusters chemistry, and consequently new questions arise on the basis of these developments \cite{gates1995supported,lopez2003synthesis,jena2006clusters,huseyinova2016synthesis,liu2021confining,luo2022topological,lee2022there}. This interest is due to an atomic clusters containing up to a few dozen atoms exhibit features that are very different from the corresponding bulk properties and that can depend very sensitively on cluster size \cite{yang2006structure}. In particular, many of these transition metal clusters are used in the field of catalysis \cite{gates1995supported,reetz1994size,aiken1999review,parkinson2017unravelling}. One of the basic principles of catalysis is that when the smaller the metal particles, the larger the fraction of the metal  atoms that are exposed at surfaces, where they are accessible to reactant molecules and available for catalysis \cite{gates1995supported}. It is well known in chemistry that the encounter between two molecules can give rise to a chemical reaction, and from the mathematical aspect there are two fundamental ways to represent these types of situations as continuous, represented by differential equations whose variables are concentrations, or as discrete, represented by stochastic processes whose variables are the number of molecules \cite{gibson2000efficient}.

Without loss of generality, it can be considered that the molecular chemisorption is due to the encounter between a molecule and a surface (or a cluster in this case) with the energy necessary for the phenomenon of adsorption to occur \cite{brown1998isolation}. Besides, the kinetics of hydrogen chemisorption by neutral gas-phase metal clusters exhibits a complex dependence on both cluster size and metal type \cite{zakin1988dependence}. For different chemical purposes, for example, in the case of copper clusters (Cu$_{n}$) is very important to have control of the chemisorption of hydrogen on these clusters, \textit{i.e.} the formation of Cu$_{n}$-H$_{2}$ species \cite{kuang2011density}.

From a reductionist point of view, the molecular chemisorption is a problem of encounter between bodies: metal clusters and reactant molecules. In our first approximation (\textit{mathematical} reduction) we will consider the problem as a problem of encounter or collisions between bodies. We are interested in proposing this strategy because we are focused to answer what is the probability of meeting between $N$ hydrogen molecules ($N$-H$_{2}$) and a fixed $M$ metallic clusters ($M$-Me$_{n}$), for a given time $t$, where the H$_{2}$ move freely in a bounded volume $V$ of $\mathbb{R}^{3}$-space. Under this assumption, we are going to consider H$_{2}$-molecules and Me$_{n}$-clusters as rigid spheres of radii $r_{1}$ and $r_{2}$, respectively. Then, it is considered that there will be a collision whenever the center-to-center distance between an H$_{2}$-molecule and a Me$_{n}$-cluster is equal to $r_{12} = r_{1} + r_{2}$ \cite{gillespie1977exact}. Also, in this context we propose the H$_{2}$-molecules follow a Brownian motion, namely: (a) it has continuous trajectories (sample paths) and (b) the increments of the paths in disjoint time intervals are independent zero mean Gaussian random variables with variance proportional to the duration of the time interval \cite{schuss2009theory}.

The pioneering work of T.D. Gillespie \cite{gillespie1977exact,gillespie1976general,gillespie1977concerning} have given rise to a large number of works that are proposed different algorithms for the calculation for numerically simulating the time evolution of a well-stirred chemically reacting system, although despite recent major improvements in the efficiency of the stochastic simulation algorithm, its drawback remains the great amount of computer time that is often required to simulate a desired amount of system time \cite{gillespie2001approximate}. While our method is a simple reduction to collisions of molecules, allows to calculate the probability of encounter (scheduled in \textbf{$R$}) for a large number of molecules ($\approx 10^{6}$) and clusters ($\approx 10^{20}$) with advantages regarding the cost of calculation, and the effects of first approximation can provide statistical support to the design of experiments. This calculation is possible using a stochastic model (Wiener process) in the context of considerations from the Maxwell-Boltzmann theory.

\section{A first theoretical approaching}
\label{sec:headings}

As we announced in the introduction, we will assume that hydrogen molecules have a random movement, whence let $\mathcal{H}(t)=(X(t),Y(t),Z(t))$ the random variable which specify the space point where the H$_{2}$ hydrogen molecule is at time $t$. Trivially, $\mathcal{H}(t)$ depends on an initially point $\mathcal{H}(0)=(x_0,y_0,z_0)$. Thus, when the initial starting point is undefined, $\mathcal{H}(t)=\mathcal{H}(t,x_0,y_0,z_0)$. Our interest is in how probably is that the distance between $\mathcal{H}(t)$ and a fixed point $(a,b,c)$ is smaller than $\epsilon$. The fixed point $(a,b,c)$ are the  coordinates for Me$_{n}$.

Let's consider the random variable $D(t)$ as the variable that measures the distance between $\mathcal{H}(t)$ and the fixed point $(a,b,c)$. Following the classical Pythagorean relationship, $D(t)= \sqrt{(X(t)-a)^2 + (Y(t)-b)^2 + (Z(t)-c)^2}$, and in general $D(t)=D(t,x_0,y_0,z_0,a,b,c)$.

Now, given a time window $[0,\tau]$, let 

\newcommand{\mllave}[2]{\left\{\begin{array}{lclcl}
		#1 \\
		#2\\
	\end{array}\right.
}

\begin{equation}
R_\tau :=\mllave{1 & \mbox{if} \displaystyle \min_{t \in [0,\tau)} \displaystyle D(t) \leq \epsilon, \\ } %
{0 & \mbox{otherwise}. }
\label{eqn:sa}
\end{equation}

So, for a fixed $t_0 >0$, we define $G(t_0)= P(D(t_0)\leq \epsilon)=\int_{0}^{\epsilon}D(s)ds$. Then, $P(R_\tau=1)=\int_{0}^{\tau} G(t) dt$.

Thus, given $\tau>0$, $R_\tau$ depends only on the initial values $(x_0,y_{0},z_{0},a,b,c)$. Now, if we have $M$-Me$_{n}$, the probability that the H$_{2}$ molecule does not meet with any of the clusters is $P(R_{\tau_1}=0,R_{\tau_2}=0,...,R_{\tau_M}=0)=p_A$, where $R_{\tau_i}, i \in \{1,...,M\}$, follows the definition given in the eq. \ref{eqn:sa}.

If $N$-H$_{2}$ molecules are in the environment, let $A_j$ the event ``the j-th hydrogen molecule meet with a metallic cluster''. Under random starting points, we are interested in $P(A_{1}^C \cap A_{2}^C \cap ... \cap A_{N}^C)=p_{A}^N$ according to the independence among the hydrogen molecules. 

\subsection{Adaptation to our context}
\label{sec:2.1}

Next, we proceed to realize the analysis according to the Brownian Motion Theory \cite{schuss2009theory}, in which the movement of the particle is independent among different axis, and we are going to assume that it follows a Wiener process \cite{leimkuhler2015molecular,leimkuhler2015numerical}. Then,

\begin{align*}
& X(t) = x_0 + W_X(t) \\
& Y(t)= y_0 + W_Y(t) \\
& Z(t)= z_0 + W_Z(t)
\end{align*}

And we will say that $W_X(t)$, $W_Y(t)$ and $W_Z(t)$ are following a Wiener processes with $\sigma=\sqrt{\frac{k_bT}{m}}$, where $k_b$ is the Boltzmann's constant, $T$ is the absolute temperature in Kelvin (K) and $m$ is the H$_{2}$'s mass in kg. That is, we are imposing a physical behavior that obeys Maxwell-Boltzmann's considerations. According this:

\begin{align*}
&X(t) \sim \mathcal{N}(x_0, \sigma^2 t) \\
&Y(t) \sim \mathcal{N}(y_0, \sigma^2 t) \\
&Z(t) \sim  \mathcal{N}(z_0, \sigma^2 t)
\end{align*}

With density function $f_X(x,t \vert x_0)$, $f_Y(y,t \vert y_0)$ and $f_Z(z,t \vert z_0)$, respectively. Under these assumptions:

\begin{align*}
& f_X(x,t \vert x_0) = \frac{1}{\sqrt{2 \pi \sigma^2 t}}\exp\left\lbrace -\frac{1}{2}\left(\frac{x - x_0}{\sigma\sqrt{t}}\right)^2\right\rbrace \\
& f_Y(y,t \vert y_0) = \frac{1}{\sqrt{2 \pi \sigma^2 t}}\exp\left\lbrace -\frac{1}{2}\left(\frac{y - y_0}{\sigma\sqrt{t}}\right)^2\right\rbrace \\
& f_Z(z,t \vert z_0) = \frac{1}{\sqrt{2 \pi \sigma^2 t}}\exp\left\lbrace -\frac{1}{2}\left(\frac{z - z_0}{\sigma\sqrt{t}}\right)^2\right\rbrace
\end{align*}

\subsubsection{Unbounded conditions}
Under unbounded conditions, as it is well known, the density of the particle position in the space for a fixed $t$ follows the expression:

\begin{align*}
f_{XYZ}&(x,y,z,t \vert x_0, y_0, z_0)=f_X(x,t \vert x_0)\cdot f_Y(y,t \vert y_0)\cdot f_Z(z,t \vert z_0) = \\
&=\frac{1}{(\sqrt{2 \pi \sigma^2 t)^3}}\exp\left\lbrace -\frac{1}{2}\left(\frac{(x - x_0)^2+(y - y_0)^2+(z-z_0)^2}{\sigma^2t}\right)\right\rbrace
\end{align*}

This function is continuous in the variables $x,y,z,t$, then is also integrable in a measurable context. Because of this fact, Fubini's theorem is aplicable. Now, calling $\nu = \sqrt{\frac{(x - x_0)^2+(y - y_0)^2+(z-z_0)^2}{2\sigma^2}}$, and integrating over the variable $t$ by substitution, results:

\begin{align*}
 f_{XYZ}(x,y,z,\tau \vert x_0, y_0, z_0) & = \frac{1}{-\nu (\sqrt{2 \pi \sigma^2})^3}\int_{0}^{\tau} \frac{-\nu}{(\sqrt{t})^3}\exp\left\lbrace -\left(\frac{\nu}{\sqrt{t}}\right)^2\right\rbrace dt \\
 & = \frac{1}{\nu (\sqrt{2 \pi \sigma^2})^3}\int_{\sqrt{\frac{\nu}{\tau}}}^{\infty}e^{-u^2}du
\end{align*} 

Remembering that the $erfc$ function \cite{magnus2013formulas} is defined by:

\begin{equation*}
erfc(z)= \frac{2}{\sqrt{\pi}}\int_{z}^{\infty}e^{-t^2}dt
\end{equation*}

we conclude: 

\begin{equation*}
f_{XYZ}(x,y,z,\tau \vert x_0, y_0, z_0) = \frac{1}{2 \pi \nu (\sqrt{2\sigma^2})^3} erfc\left(\sqrt{\frac{\nu}{\tau}}\right)
\end{equation*}

From the physical-experimental perspective that the problem is lays out, the unbounded system lacks interest, so we will proceed to study the case of the bounded system.

\subsubsection{Bounded conditions} 
We assume that the experiment takes place into a cubic recipe centered at the origin. This implies that $X(t)$, $Y(t)$ and $Z(t)$ $\in [-L; L]$, for a fixed volume $V = L^{3}$ in $\mathbb{R}^{3}$-space.

In a similar issue the traditional way of approaching is by ``truncation" \cite{heckman1976common,stein1981estimation}. A drawback of this approach is the fact that the truncation does not represent precisely the reflection on the boundaries. An illustrative and motivational argument is given by the following example: suppose a random walk of $N=4$ steps, with starting point at the origin. Then, the walker moves $1$ step at right or left (with equal probability) at each step. Then, after four steps, the resultant probabilities of the walker position are:

\begin{align*}
0&;\mbox{ with probability } 3/8, \\
-2 \mbox{ or } 2&; \mbox{ with probability } 2/8, \\
-4 \mbox{ or } 4&; \mbox{ with probability } 1/16. \\
\end{align*}

The probability values (under truncation) in the closed interval $[-2,2]$ for the values $(-2,-1,0,1,2)$ are, respectively: \[(2/7,0,3/7,0,2/7)\]

With fixed boundaries, considering reflections at $[-2,2]$, we can construct the following Markov transition matrix $P$:

\begin{center}
	$P=$
	\vline
	\begin{tabular}{ccccc}
		0 &1 &0 &0 &0\\
		1/2 &0 &1/2 &0 &0\\ 
		0 &1/2 &0 &1/2 &0\\
		0 &0 &1/2 &0 &1/2\\
		0 &0 &0 &1 &0\\
	\end{tabular}\vline
\end{center}

At the fourth step, after some algebra, we obtain the respectively mass point probability for the position of the walker. This is provided by the stochastic vector \[(1/4,0,1/2,0,1/4)\] (given by the third file of $P^4$, \textit{i.e.}: with starting point at the origin). At this point, is clearly the difference between truncation and ``rebounds" (considering reflection on the boundary).

\begin{figure}
  \centering
  \includegraphics[width=10cm]{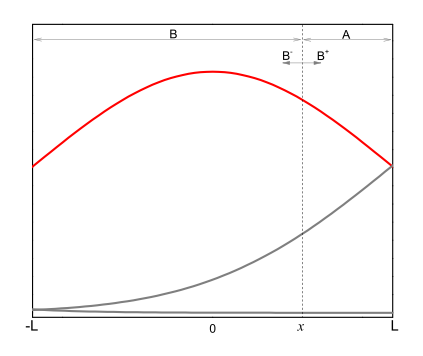}
  \caption{In red colour an arbitrary normal distribution, $\mathcal{N}(0, \sigma)$. We observe in gray colour the folding of the normal distribution at the edge of the box. We could see that $A+B$ = $2L$.}
  \label{fig:fig1}
\end{figure}

We must modify the density of the position $\mathcal{H}(t)$ according to the particle rebounds (see Fig. \ref{fig:fig1}). It is important to note that the rebounds indicated in the figure in gray colour do not correspond to the physical rebounds of the particles in the cubic box, but to the contributions of the displaced distribution considering an infinite behavior.

Inside the box, the derived density $f_B$ according to the variable $X(t) \sim \mathcal{N}(x_0,\sigma^2t)$ with density function $f_X$ of the particle position (for each dimension, see Fig. \ref{fig:fig1}) follows the expression:

\begin{equation*}
f_B(x) = \left[f_X(x) + f_{B^+}(x) + f_{B^-}(x)\right]\times I_{[-L; L]}(x)
\end{equation*}

where:

\begin{align*}
f_{B^+}(x) &= f(x + 2 A) + f(x + 2 A + 2 B) + f(x + 2 A + 2 B + 2 A) + ... = \\
&= f(x +2(L-x)) + f(x +2(L-x) + 2(x-(-L)) + ...   \\
&= f(-x + 2L) + f(x + 4L) + f(-x + 6L) + ...   \\
&= \sum_{k=1}^{\infty}f((-1)^k x + 2kL)  \\
&= \sum_{k=1}^{\infty} \frac{1}{\sqrt{2 \pi \sigma^2 t}}\exp\left\lbrace -\frac{1}{2} \left(\frac{((-1)^k x + 2kL)-x_0}{\sigma \sqrt{t}}\right)^2\right\rbrace
\end{align*}

and

\begin{align*}
f_{B^-}(x) &= f(x - 2 B) + f(x - 2 B - 2 A) + f(x - 2 B - 2 A - 2 B) + ... = \\
&= f(x -2(x-(-L))) + f(x -2(x-(-L)) - 2(L-x)) + ...   \\
&= f(-x - 2L) + f(x - 4L) + f(-x - 6L) + ...   \\
&= \sum_{k=1}^{\infty}f((-1)^k x - 2kL)  \\
&= \sum_{k=1}^{\infty} \frac{1}{\sqrt{2 \pi \sigma^2 t}}\exp\left\lbrace -\frac{1}{2} \left(\frac{((-1)^k x - 2kL)-x_0}{\sigma \sqrt{t}}\right)^2\right\rbrace
\end{align*}

The proof that $f_B$ is a density function is straightforward its definition. Trivially, $f_B>0$, and by construction:
\[ \int_{-\infty}^{\infty}f_B(t)dt=\int_{-L}^{L}f_B(t)dt=\int_{-\infty}^{\infty}f_X(t)dt=1\] 

For practical purposes, we now try to find an upper bound to this expression. Looking at in the model proposed, the next constraint is straightforward $\vert (-1)^k x -x_0 \vert \leq 2L$.

Following these constraints:

\begin{align*}
f_{B^+}(x) &= \sum_{k=1}^{\infty} \frac{1}{\sqrt{2 \pi \sigma^2 t}}\exp\left\lbrace -\frac{1}{2} \left(\frac{((-1)^k x + 2kL)-x_0}{\sigma \sqrt{t}}\right)^2\right\rbrace  \\
&\leq \sum_{k=1}^{\infty} \frac{1}{\sqrt{2 \pi \sigma^2 t}}\exp\left\lbrace -\frac{1}{2} \left(\frac{-2L + 2kL}{\sigma \sqrt{t}}\right)^2\right\rbrace  \\
&= \sum_{k=1}^{\infty} \frac{1}{\sqrt{2 \pi \sigma^2 t}}\exp\left\lbrace -\frac{1}{2} \left(\frac{2(k-1)L}{\sigma \sqrt{t}}\right)^2\right\rbrace   \\
&= \sum_{k=0}^{\infty} \frac{1}{\sqrt{2 \pi \sigma^2 t}}\exp\left\lbrace -\frac{1}{2} \left(\frac{2kL}{\sigma \sqrt{t}}\right)^2\right\rbrace  \\
&= \frac{1}{\sqrt{2 \pi \sigma^2 t}} \sum_{k=0}^{\infty} \exp\left\lbrace -\frac{1}{2} \left(\frac{4L^2}{\sigma^2 t}\right)\right\rbrace^{k^2} \\
\end{align*}

It is known that $\displaystyle \sum_{k=0}^{\infty}\ r^{k^{2}} = \frac{1}{2} + \frac{1}{2}\Theta_{_{E}}[3, 0, r]$  where $\Theta_{_{E}}$ is the Jacobi theta elliptic function \cite{magnus2013formulas}. So:
\begin{align*}
	f_{B^+}(x)& \leq \frac{1}{\sqrt{2 \pi \sigma^2 t}} \sum_{k=0}^{\infty} \exp\left\lbrace -\frac{1}{2} \left(\frac{4L^2}{\sigma^2 t}\right)\right\rbrace^{k^2} \\
	&=\frac{1}{\sqrt{2 \pi \sigma^2 t}} \left(\frac{1}{2} + \frac{1}{2}\Theta_{_{E}} \left[ 3, 0, exp\left\lbrace -\frac{1}{2} \left(\frac{4L^2}{\sigma^2 t}\right)\right\rbrace \right]\right)
	\end{align*}
 
and

	\begin{align*}
	f_{B^-}(x) &= \sum_{k=1}^{\infty} \frac{1}{\sqrt{2 \pi \sigma^2 t}}\exp\left\lbrace -\frac{1}{2} \left(\frac{((-1)^k x - 2kL)-x_0}{\sigma \sqrt{t}}\right)^2\right\rbrace  \\
	&\leq \sum_{k=1}^{\infty} \frac{1}{\sqrt{2 \pi \sigma^2 t}}\exp\left\lbrace -\frac{1}{2} \left(\frac{-2L - 2kL}{\sigma \sqrt{t}}\right)^2\right\rbrace  \\
	&= \sum_{k=1}^{\infty} \frac{1}{\sqrt{2 \pi \sigma^2 t}}\exp\left\lbrace -\frac{1}{2} \left(\frac{-2(k+1)L}{\sigma \sqrt{t}}\right)^2\right\rbrace   \\
	&= \sum_{k=2}^{\infty} \frac{1}{\sqrt{2 \pi \sigma^2 t}}\exp\left\lbrace -\frac{1}{2} \left(\frac{-2kL}{\sigma \sqrt{t}}\right)^2\right\rbrace  \\
		&= \frac{1}{\sqrt{2 \pi \sigma^2 t}} \left(\sum_{k=0}^{\infty} \exp\left\lbrace -\frac{1}{2} \left(\frac{4L^2}{\sigma^2 t}\right)\right\rbrace^{k^2} - 1 - \exp\left\lbrace -\frac{1}{2} \left(\frac{4L^2}{\sigma^2 t}\right)\right\rbrace\right)\\
	&= \frac{1}{\sqrt{2 \pi \sigma^2 t}} \left(\frac{1}{2} + \frac{1}{2}\Theta_{_{E}} \left[ 3, 0, exp\left\lbrace -\frac{1}{2} \left(\frac{4L^2}{\sigma^2 t}\right)\right\rbrace \right]- 1 - \exp\left\lbrace -\frac{1}{2} \left(\frac{4L^2}{\sigma^2 t}\right)\right\rbrace\right)
	\end{align*}

 Then,
 
\begin{align*}
f_{B^+}(x) + f_{B^-}(x) & \leq \frac{1}{\sqrt{2 \pi \sigma^2 t}} \left(\Theta_{_{E}} \left[ 3, 0, exp\left\lbrace -\frac{1}{2} \left(\frac{4L^2}{\sigma^2 t}\right)\right\rbrace \right] - \exp\left\lbrace -\frac{1}{2} \left(\frac{4L^2}{\sigma^2 t}\right)\right\rbrace\right)\\ & = C_B
	\end{align*}

For each $x \in [-L,L]$, $f_B(x) \leq f_X(x) + C_B$. Besides $C_B$ does not depends on $x$. Consequently, we have a maximum for the density $f_B$ which is equal to $f(x_0) + C_B$.

Calling $P_B=(f(x_0) + C_B).2\epsilon$, we can conclude that:
	\[P(X(t)\in (a_0 - \epsilon, a_0 + \epsilon))\leq P_B, \text{for any $a_0$.}\]
 
Analogous, $C_B$ is the same for the variables $Y(t)$ and $Z(t)$, and we know that $f(x_0)=f(y_0)=f(z_0)$. Then, the same result is available for the variables $Y(t)$ and $Z(t)$. According the bounded $C_B$, it is straightforward the uniform convergence of the series $f_{B^+}$ and $f_{B^-}$ (by the M Weierstrass criteria). An important fact to remark is that $P_B$ is not even a probability, but in the case in we are  interested, we know that is a real number bigger than the probability desired, and then, under certain conditions, we can work with it.

For practical purposes, the error through the $C_B$ implementation can be minimized, since the first $S$ terms are available, and the tail can be compared with

\begin{align*}
\sum_{k=0}^{S-1}r^{k^2} & \leq \sum_{k=0}^{\infty}r^{k^2} = \sum_{k=0}^{S-1}r^{k^2} + \sum_{k=S}^{\infty}r^{k^2} 
\end{align*}

And,

\[ \sum_{k=S}^{\infty}r^{k^2} = \sum_{k=0}^{\infty}r^{(k+S)^2} =\sum_{k=0}^{\infty}r^{k^2+2kS+S^2}=r^{S^2}\sum_{k=0}^{\infty}r^{k^2}r^{2kS}\leq r^{S^2}\sum_{k=0}^{\infty}r^{k^2}\]

Then,

\begin{align*}
\sum_{k=S}^{\infty}r^{k^2} & \leq r^{S^2} \left( \frac{1}{2} + \frac{1}{2}\Theta_{_{E}}[3, 0, r]\right) 
\end{align*}

Controlling the value of S controls the value of the error made by truncating the sum. As we said, $C_B$ does not depend on $x$, thus, the desired probability can be estimated with any degree of accuracy, according the computational cost necessary to this development.

Taking into consideration the Brownian Motion Theory, in the time lapse of 1 second, the particle position under unbounded conditions follows a $\mathcal{N}(x_0,\sigma^2)$ distribution. To discretize the problem, if we partitioned the time axis of $\tau$ seconds in $\tau$ intervals of 1 second each one, then:

\begin{align*}
 P(\mathcal{H}(t) \in B_{\epsilon}(a,b,c)) \leq P(\mathcal{H}(t) \in Q_{\epsilon}(a,b,c))
\end{align*}

where $Q_{\epsilon}(a,b,c)$ denotes the cube centered in $(a,b,c)$ with side size $2 \times \epsilon$. And, considering the independence between $X(t)$, $Y(t)$ and $Z(t)$, with $X(t)\in (a-\epsilon,a+\epsilon)$, $Y(t)\in (b-\epsilon,b+\epsilon)$ and $Z(t)\in (c-\epsilon,c+\epsilon)$, $P( \mathcal{H}(t) \in Q_{\epsilon}(a,b,c)) = P( \mathcal{H} \in Q_{\epsilon})$
is

\begin{align*}
P( \mathcal{H} \in Q_{\epsilon}) = P(X(t))\times P(Y(t))\times P(Z(t)) \leq P_B\times P_B\times P_B = P_B^3
\end{align*}

For each second $\tau_j$ for $\tau_j \in \{1:\tau\}$, $P(\mathcal{H}(t)\in Q_{\epsilon}(a,b,c)) \leq P_B^3$. Then, under the Wiener process formulation, $\mathcal{H}(\tau_j) \perp \mathcal{H}(\tau_k\vert \tau_j)$ if $j \neq k, j\leq k$.

$P(\mathcal{H}(\tau_j)\in Q_\epsilon(a,b,c)) \leq P_B^3, \forall \tau_j \in \{1:\tau\}$. Calling $F:=$``$\#$ of $\tau_j \in \{1:T\}$ in which $H(\tau_j) \in Q_{\epsilon}(a,b,c)$", we are interesting in the event $F=0$.

According its nature, $F$ is a Binomial random variable $\mathcal{B}(\tau,P_B^3)$. Consequently, the \textit{non-collision} probability is $p_{_{NC}} = P(F=0) \leq (1 - P_B^3)^\tau$. At this point, we only can conclude that the probability of the encounter between a hydrogen molecule and a Me$_n$ cluster in a time $\tau$ is less than $p$. We proceed to analyze what happens when the number of hydrogen molecules and metallic clusters increase. We emphasize that the H$_{_{2}}$ molecules have a random movement while the clusters are confined in a fixed region of space.
Since $p$ is the probability that a random hydrogen molecule meets in the cube $Q_\epsilon$ in which a Me$_n$ cluster is, the most unfavorable case with $M$ clusters is when there is no intersection among the cubes that contain it. In this case:

\begin{align*}
p_A &=P(R_{\tau_1}=0,...,R_{\tau_M}=0) \\ 
&=1-\bigcup_{i=1}^{M}P(R_{\tau_i}=1) \\
&\geq 1-\left(\sum_{i=1}^{M}P(R_{\tau_i}=1)\right) \\
&=1-M\times p
\end{align*}

In view of this analysis, we can conclude that the \textit{non-collision} probability is higher than $p_{NC}$.

In regular conditions, when this approach is used, the values of $p_{NC}$ and $N$ outcomes into a several numerical instability. In this case, the small value of $p_{NC}$ and the large value of $N$ place us in conditions to use the Poisson approach to the Binomial distribution (with parameter $\lambda = N \times p$). Then, $P(X=0)\approx \exp(-\lambda)$. Even in the cases when the probability is still unavailable, the expected number of collisions is presented according a time window, and then we can estimate the probability of collisions in a time window $T$ using the relationship between the Poisson and Exponential distributions\cite{gerritsen1977encounter}.

Next, we present the results of the analysis whit different box dimensions (in meters) and number of hydrogen molecules ($N$), according to $M=1.9 \times 10^{^{20}}$ Cu$_{_{20}}$-clusters \cite{andrini2019structure}, where the Cu$_{_{20}}$-clusters have been considered as spheres.

\section{Results and analysis}

\subsection{Obtaining \textit{non-collision} probability values}

The situation we consider is approximately a \textquotedblleft realistic\textquotedblright situation, with $M=1.9 \times 10^{^{20}}$  Cu$_{_{20}}$-clusters in a cubic box according to the standard dimensions of reaction chambers (0.1$^{^{3}}$, 0.2$^{^{3}}$ and 0.4$^{^{3}}$ m$^{^{3}}$), and a variable $N$-H$_{_{2}}$-molecules \textquotedblleft contamination\textquotedblright ($10^{^{1}} \leq N \leq 10^{^{10}}$). It worked with three temperatures, T, 293, 373 and 473 K. The choice of T is arbitrary, conditioned by the possible reaction temperatures \cite{corma2013exceptional}.

\begin{figure}
\centering
	\includegraphics[width=10cm]{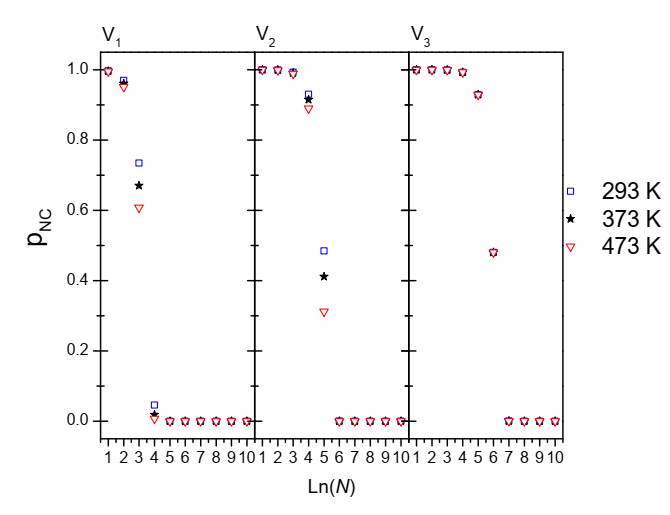}
	\caption{Results for the \textit{non-collision} probability, $p_{_{NC}}$, \textit{vs.} $\ln (N)$ for $L=0.05 m$ (V$_{_{1}}$), $L=0.1 m$ (V$_{_{2}}$) and $L=0.2 m$ (V$_{_{3}}$), at T = 293 K (blue square), 373 K (black star) and 493 K (red triangle).}
	\label{fig:fig2}
\end{figure}

In Fig. \ref{fig:fig2} we observe the results obtained for the simulations, considering the maximum sum. That is, take $S=10^{^{6}}$, perform the sum, and add the maximum level for the error. Clearly, a greater probability of non-collision, $p_{_{NC}}$, is observed depending on the increase in volume.

For a detailed study, we proceed as follows: we model the data obtained through a non-linear graphic fitting considering a Boltzmann decrease function, $g(x)=A_{_{2}}+\frac{A_{_{1}}-A_{_{2}}}{1+\exp \left( \frac{x-x_{_{0}}}{dx}\right) }$ (see Fig. \ref{fig:fig3}). In the Appendix A.2 we show the statistical results for each parameter in each data fitting.

Under these considerations, we can calculate the critical value (criticality)\cite{bak1995complexity,Williams1992} of hydrogen molecules, that is \textquotedblleft what is the value of N for which the  \textit{non-collision} probability is greater than $\frac{1}{2}$\textquotedblright, \textit{i.e.} the value of the exponent for which $\frac{1}{2}< p_{_{NC}}$.

It should be clarified that, in the strict physical sense, there is no abrupt phase transition to consider \textquotedblleft criticality\textquotedblright. As we assumed in the introduction, we consider that there is a chemical reaction if there is an encounter between two molecules, and under this assumption we are considering as critical the level of presence of hydrogen for a chemical reaction to occur. In any case, it can be demonstrated that there is an \textquotedblleft abrupt\textquotedblright transition behavior, for a well defined interval in the number of molecules. In Fig. \ref{fig:fig3} we can observe this behavior.

\begin{figure}
\centering
	\includegraphics[width=10cm]{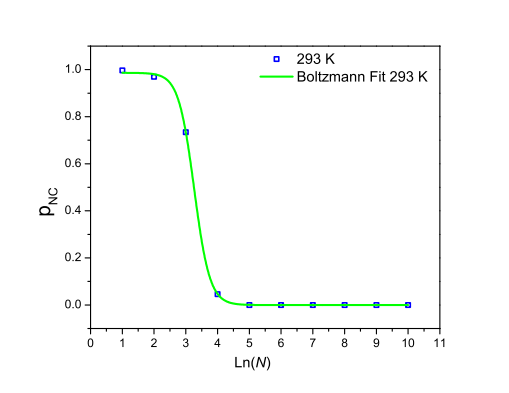}
	\caption{Data (blue square) modeling using a non-linear Boltzmann decrease function (green line).}
	\label{fig:fig3}
\end{figure}

In Table \ref{tab:table1} we can see the critical values obtained from the decrease model for each box and each temperature. For the smallest volumes, V$_{_{1}}$ and  V$_{_{2}}$, it is observed that the critical value of $N$ depends more strongly on the temperature than in the case of the larger volume (V$_{_{3}}$). Although it is remarkable the fact of dependence with the size of the box, it can be seen directly from Fig. \ref{fig:fig2}. In this way, and under these simplified assumptions, we can obtain control of contaminant molecules in relation to the volume and temperature parameters. Linear behavior is evident from the values obtained (Table \ref{tab:table1}, $N$ \textit{vs.} temperature). Moreover, as the volume increases the slope increases from negative values to null value.

\begin{table}
\caption{Critical values obtained from the decrease model for each box and each temperature, for \textit{mathematical robounds}.}
\centering
\begin{tabular}{llll}
\hline\noalign{\smallskip}
L [m] & 293 K & 373 K & 473 K \\
\noalign{\smallskip}\hline\noalign{\smallskip}
0.05 & 3.25 & 3.16 & 3.09 \\
0.10 & 4.97 & 4.86 & 4.72 \\
0.20 & 5.96 & 5.96 & 5.96 \\
\noalign{\smallskip}\hline
\end{tabular}
\label{tab:table1}
\end{table}

\begin{figure}
\centering
	\includegraphics[width=10cm]{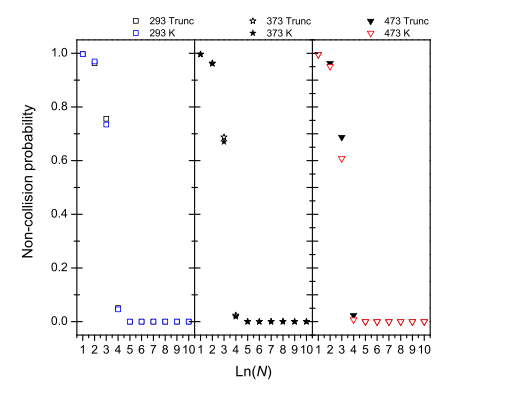}
	\caption{Results for the \textit{non-collision} probability, $p_{_{NC}}$, \textit{vs.} $\ln (N)$ for $L=0.05 m$ (V$_{_{1}}$), at T = 293 K, 373 K and 493 K. Comparison between models:\textquotedblleft xxx Trunc\textquotedblright correspond to the truncated normal model and \textquotedblleft xxx K\textquotedblright to the \textit{mathematical rebound} model.}
	\label{fig:fig4}    
\end{figure}

\begin{figure}
\centering
	\includegraphics[width=10cm]{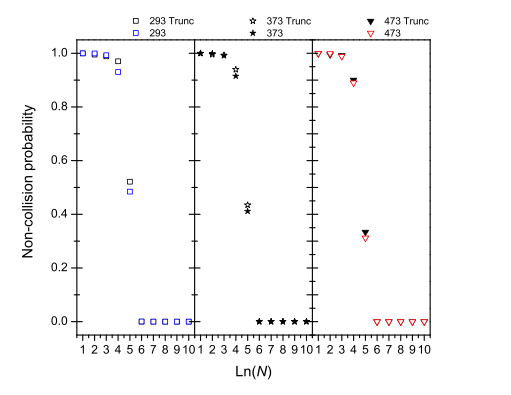}
	\caption{Results for the \textit{non-collision} probability, $p_{_{NC}}$, \textit{vs.} $\ln (N)$ for $L=0.1 m$ (V$_{_{1}}$), at T = 293 K, 373 K and 493 K. Comparison between models:\textquotedblleft xxx Trunc\textquotedblright correspond to the truncated normal model and \textquotedblleft xxx K\textquotedblright to the \textit{mathematical rebound} model.}
	\label{fig:fig5}
\end{figure}

\begin{figure}
\centering
	\includegraphics[width=10cm]{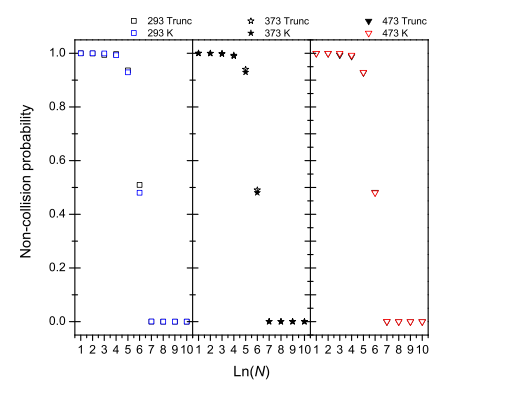}
	\caption{Results for the \textit{non-collision} probability, $p_{_{NC}}$, \textit{vs.} $\ln (N)$ for $L=0.2 m$ (V$_{_{1}}$), at T = 293 K, 373 K and 493 K. Comparison between models:\textquotedblleft xxx Trunc\textquotedblright correspond to the truncated normal model and \textquotedblleft xxx K\textquotedblright to the \textit{mathematical rebound} model.}
	\label{fig:fig6}
\end{figure}

\begin{table}
	\caption{Critical values obtained from the decrease model for each box and each temperature, for truncated normal model.}
    \centering
	\begin{tabular}{llll}
		\hline\noalign{\smallskip}
		L [m] & 293 K & 373 K & 473 K \\
		\noalign{\smallskip}\hline\noalign{\smallskip}
		0.05 & 3.27 & 3.18 & 3.12 \\
		0.10 & 5.01 & 4.91 & 4.76 \\
		0.20 & 6.00 & 5.98 & 5.96 \\
		\noalign{\smallskip}\hline
	\end{tabular}
 \label{tab:table2}
\end{table}

\section{Conclusion}
By way of conclusion, it can be indicated that considering a Wiener stochastic process, for thermodynamic-statistical movements of a gas confined in a box, and considering \textit{mathematical rebounds} bounded by the physical-geometric contour of the problem, the analytical expression could be obtained for the probability density function of encounters between two differentiated species of molecules (one of the species fixed in the box -solid or liquid- and the other species is a gas whose molecules move stochastically). In addition, the function obtained can be calculated numerically or can be bounded. The bounded process allows to reduce the computational cost, and to limit the error from cutting the sum in a finite number. In particular, there is an error control that can be made, and it is possible to refine the process according to the precision required.

From the physical-chemical point of view, it is observed that both the number of gas molecules and the dimensions of the box affect the probability of encounter. For this model, temperature is a parameter that has a lower incidence on the values of the probability of encounter. At this point some considerations have to be made. The first is that in a strict sense a chemical reaction is more than the encounter of two chemical entities. The second is the exceptional chemical nature of metal clusters, which make them highly reactive. Despite the simplicity of the model we are proposing, this model can account in an experiment design about the collision probability between two chemical entities (and this collision can lead to a chemical reaction).

From the point of view of computation, it is a system that requires less computational cost (time + memory) than the algorithmic systems developed for this type of problems, so it contributes as a test method in the design of experiments.

The comparison with an established method (truncated normal model) was optimal. In the method of mathematical rebounds the number of molecules needed for a reaction is less than the number obtained by the truncated normal model. This is an advantage when strict contamination control is needed.

On the other hand, in terms of obtaining the density function, mathematical results can be generalized for volumes of rectangular prisms of uneven sides. In addition, it remains to calculate the first and second order moments of the density function obtained, work that exceeded the purposes of present communication.

\section*{Acknowledgments}
This was was supported in part by PICT-2019-0784, PICT-2017-3944, PICT-2017-1220, PICT-2017-3150 (PICT, Agencia Nacional de Promoción de la Investigación, el Desarrollo Tecnológico y la Innovación) and PPID-I231 (PPID, Universiad Nacional de La Plata).

\bibliographystyle{unsrt}  
\bibliography{references}  

\begin{thebibliography}{10}

\bibitem{gates1995supported}
Bruce~C. Gates.
\newblock Supported metal clusters: synthesis, structure, and catalysis.
\newblock {\em Chemical reviews}, 95(3):511--522, 1995.

\bibitem{lopez2003synthesis}
M~Arturo L\'opez-Quintela.
\newblock Synthesis of nanomaterials in microemulsions: formation mechanisms
  and growth control.
\newblock {\em Current Opinion in Colloid \& Interface Science}, 8(2):137--144,
  2003.

\bibitem{jena2006clusters}
Puru Jena and A.~Welford Castleman~Jr.
\newblock Clusters: A bridge across the disciplines of physics and chemistry.
\newblock {\em Proceedings of the National Academy of Sciences},
  103(28):10560--10569, 2006.

\bibitem{huseyinova2016synthesis}
Shahana Huseyinova, Jose\'e Blanco, Fe\'elix~G. Requejo, Jose\'e~M
  Ramallo-L\'opez, M~Carmen Blanco, David Buceta, and M~Arturo
  Lo\'opez-Quintela.
\newblock Synthesis of highly stable surfactant-free cu5 clusters in water.
\newblock {\em The Journal of Physical Chemistry C}, 120(29):15902--15908,
  2016.

\bibitem{liu2021confining}
Lichen Liu and Avelino Corma.
\newblock Confining isolated atoms and clusters in crystalline porous materials
  for catalysis.
\newblock {\em Nature Reviews Materials}, 6(3):244--263, 2021.

\bibitem{luo2022topological}
Huixia Luo, Peifeng Yu, Guowei Li, and Kai Yan.
\newblock Topological quantum materials for energy conversion and storage.
\newblock {\em Nature Reviews Physics}, 4(9):611--624, 2022.

\bibitem{lee2022there}
Seunghoon Lee, Joonho Lee, Huanchen Zhai, Yu~Tong, Alexander~M Dalzell,
  Ashutosh Kumar, Phillip Helms, Johnnie Gray, Zhi-Hao Cui, Wenyuan Liu, et~al.
\newblock Is there evidence for exponential quantum advantage in quantum
  chemistry?
\newblock {\em arXiv preprint arXiv:2208.02199}, 2022.

\bibitem{yang2006structure}
Mingli Yang, Koblar~A Jackson, Christof Koehler, Thomas Frauenheim, and Julius
  Jellinek.
\newblock Structure and shape variations in intermediate-size copper clusters.
\newblock {\em The Journal of chemical physics}, 124(2):024308, 2006.

\bibitem{reetz1994size}
Manfred~T Reetz and Wolfgang Helbig.
\newblock Size-selective synthesis of nanostructured transition metal clusters.
\newblock {\em Journal of the American Chemical Society}, 116(16):7401--7402,
  1994.

\bibitem{aiken1999review}
John~D Aiken~III and Richard~G Finke.
\newblock A review of modern transition-metal nanoclusters: their synthesis,
  characterization, and applications in catalysis.
\newblock {\em Journal of Molecular Catalysis A: Chemical}, 145(1-2):1--44,
  1999.

\bibitem{parkinson2017unravelling}
Gareth~S Parkinson.
\newblock Unravelling single atom catalysis: The surface science approach.
\newblock {\em arXiv preprint arXiv:1706.09473}, 2017.

\bibitem{gibson2000efficient}
Michael~A Gibson and Jehoshua Bruck.
\newblock Efficient exact stochastic simulation of chemical systems with many
  species and many channels.
\newblock {\em The journal of physical chemistry A}, 104(9):1876--1889, 2000.

\bibitem{brown1998isolation}
David~E Brown, Douglas~J Moffatt, and Robert~A Wolkow.
\newblock Isolation of an intrinsic precursor to molecular chemisorption.
\newblock {\em Science}, 279(5350):542--544, 1998.

\bibitem{zakin1988dependence}
MR~Zakin, RO~Brickman, DM~Cox, and A~Kaldor.
\newblock Dependence of metal cluster reaction kinetics on charge state. ii.
  chemisorption of hydrogen by neutral and positively charged iron clusters.
\newblock {\em The Journal of chemical physics}, 88(10):6605--6610, 1988.

\bibitem{kuang2011density}
Xiang-Jun Kuang, Xin-Qiang Wang, and Gao-Bin Liu.
\newblock A density functional study on the adsorption of hydrogen molecule
  onto small copper clusters.
\newblock {\em Journal of Chemical Sciences}, 123(5):743--754, 2011.

\bibitem{gillespie1977exact}
Daniel~T Gillespie.
\newblock Exact stochastic simulation of coupled chemical reactions.
\newblock {\em The journal of physical chemistry}, 81(25):2340--2361, 1977.

\bibitem{schuss2009theory}
Zeev Schuss.
\newblock {\em Theory and applications of stochastic processes: an analytical
  approach}, volume 170.
\newblock Springer Science \& Business Media, 2009.

\bibitem{gillespie1976general}
Daniel~T Gillespie.
\newblock A general method for numerically simulating the stochastic time
  evolution of coupled chemical reactions.
\newblock {\em Journal of computational physics}, 22(4):403--434, 1976.

\bibitem{gillespie1977concerning}
Daniel~T Gillespie.
\newblock Concerning the validity of the stochastic approach to chemical
  kinetics.
\newblock {\em Journal of Statistical Physics}, 16(3):311--318, 1977.

\bibitem{gillespie2001approximate}
Daniel~T Gillespie.
\newblock Approximate accelerated stochastic simulation of chemically reacting
  systems.
\newblock {\em The Journal of chemical physics}, 115(4):1716--1733, 2001.

\bibitem{leimkuhler2015molecular}
Ben Leimkuhler and Charles Matthews.
\newblock Molecular dynamics.
\newblock {\em Interdisciplinary applied mathematics}, 36, 2015.

\bibitem{leimkuhler2015numerical}
Ben Leimkuhler and Charles Matthews.
\newblock Numerical methods for stochastic molecular dynamics.
\newblock In {\em Molecular Dynamics}, pages 261--328. Springer, 2015.

\bibitem{magnus2013formulas}
Wilhelm Magnus, Fritz Oberhettinger, and Raj~Pal Soni.
\newblock {\em Formulas and theorems for the special functions of mathematical
  physics}, volume~52.
\newblock Springer Science \& Business Media, 2013.

\bibitem{heckman1976common}
James~J Heckman.
\newblock The common structure of statistical models of truncation, sample
  selection and limited dependent variables and a simple estimator for such
  models.
\newblock In {\em Annals of economic and social measurement, volume 5, number
  4}, pages 475--492. NBER, 1976.

\bibitem{stein1981estimation}
Charles~M Stein.
\newblock Estimation of the mean of a multivariate normal distribution.
\newblock {\em The annals of Statistics}, pages 1135--1151, 1981.

\bibitem{gerritsen1977encounter}
Jeroen Gerritsen and J~Rudi Strickler.
\newblock Encounter probabilities and community structure in zooplankton: a
  mathematical model.
\newblock {\em Journal of the Fisheries Board of Canada}, 34(1):73--82, 1977.

\bibitem{andrini2019structure}
Leandro Andrini, Germ{\'a}n~J Soldano, Marcelo~M Mariscal, F{\'e}lix~G Requejo,
  and Yves Joly.
\newblock Structure stability of free copper nanoclusters: Fsa-dft cu-building
  and fdm-xanes study.
\newblock {\em Journal of Electron Spectroscopy and Related Phenomena},
  235:1--7, 2019.

\bibitem{corma2013exceptional}
Avelino Corma, Patricia Concepci{\'o}n, Mercedes Boronat, Mar{\'i}a~J Sabater,
  Javier Navas, Miguel~Jos{\'e} Yacaman, Eduardo Larios, {\'A}lvaro Posadas,
  M~Arturo L{\'o}pez-Quintela, David Buceta, Ernest Mendoza, Gemma Guilera, and
  {\'A}lvaro Mayoral.
\newblock Exceptional oxidation activity with size-controlled supported gold
  clusters of low atomicity.
\newblock {\em Nature Chemistry}, 5(9):775--781, 2013.

\bibitem{bak1995complexity}
Per Bak and Maya Paczuski.
\newblock Complexity, contingency, and criticality.
\newblock {\em Proceedings of the National Academy of Sciences},
  92(15):6689--6696, 1995.

\bibitem{Williams1992}
Terrie~M. Williams.
\newblock Criticality in stochastic networks.
\newblock {\em Journal of the Operational Research Society}, 43(4):353--357,
  1992.

\end{thebibliography}

\vspace{1 cm}

\textbf{Appendix}

\vspace{0.3 cm}

\textbf{A.1}

Errors in the Boltzmann model for the probability calculated according to \textit{mathematical rebounds}.

Program used: Origin 9.1

In all cases, number of points is 10, and degrees of freedon is 6.
\vspace{0.5 cm}

L=0.05 m, T = 293 K
\begin{tabular}{ccc}
	Parameter & Value &	Standard Error \\
	A1 &	0.991 &	0.009 \\
	A2 &	-0.0060 &	0.0008 \\
	x0 &	4.98 &	0.02 \\
	dx &	0.30 &	0.03 \\
\end{tabular}		

Reduced Chi-Sqr	2.66387 $\times$ 10$^{^{-4}}$ \\
Residual Sum of Squares: 0.0016 \\
Adj. R-Square:	0.99888 \\
\vspace{0.5 cm}

L=0.05 m, T = 373 K
\begin{tabular}{ccc}
	Parameter & Value &	Standard Error \\
	A1 &	0.981 &	0.006 \\
	A2 &	-0.005 &	0.003 \\
	x0 &	3.16 &	0.01 \\
	dx &	0.22 &	0.02 \\
\end{tabular}

Reduced Chi-Sqr	7.93743 $\times$ 10$^{^{-5}}$ \\
Residual Sum of Squares: 4.76246 $\times$ 10$^{^{-4}}$ \\
Adj. R-Square:	0.99957 \\
\vspace{0.5 cm}

L=0.05 m, T = 473 K
\begin{tabular}{ccc}
	Parameter & Value &	Standard Error \\
	A1 &	0.976 &	0.008 \\
	A2 &	-0.0011 &	0.0009 \\
	x0 &	3.10 &	0.02 \\
	dx &	0.21 &	0.03 \\
\end{tabular}	

Reduced Chi-Sqr	1.30175 $\times$ 10$^{^{-4}}$ \\
Residual Sum of Squares: 7.8105 $\times$ 10$^{^{-4}}$ \\
Adj. R-Square:	0.99927 \\
\vspace{0.5 cm}

L=0.1 m, T = 293 K
\begin{tabular}{ccc}
	Parameter & Value &	Standard Error \\
	A1 &	0.991 &	0.009 \\
	A2 &	-0.0060 &	0.0011 \\
	x0 &	4.98 &	0.02 \\
	dx &	0.30 &	0.03 \\
\end{tabular}	

Reduced Chi-Sqr	2.66387 $\times$ 10$^{^{-4}}$ \\
Residual Sum of Squares: 0.0016 \\
Adj. R-Square:	0.99888 \\
\vspace{0.5 cm}

L=0.1 m, T = 373 K
\begin{tabular}{ccc}
	Parameter & Value &	Standard Error \\
	A1 &	0.994 &	0.008 \\
	A2 &	-0.006 &	0.002 \\
	x0 &	4.88 &	0.02 \\
	dx &	0.33 &	0.02 \\
\end{tabular}

Reduced Chi-Sqr	1.88888 $\times$ 10$^{^{-4}}$ \\
Residual Sum of Squares: 0.00113 \\
Adj. R-Square:	0.9992 \\
\vspace{0.5 cm}

L=0.1 m, T = 473 K
\begin{tabular}{ccc}
	Parameter & Value &	Standard Error \\
	A1 &	0.996 &	0.005 \\
	A2 &	-0.0043 &	0.0019 \\
	x0 &	4.73 &	0.01 \\
	dx &	0.33 &	0.01 \\
\end{tabular}	

Reduced Chi-Sqr	7.76599 $\times$ 10$^{^{-5}}$ \\
Residual Sum of Squares: 4.65959 $\times$ 10$^{^{-4}}$ \\
Adj. R-Square:	0.99967 \\
\vspace{0.5 cm}

L=0.2 m, T = 293 K
\begin{tabular}{ccc}
	Parameter & Value &	Standard Error \\
	A1 &	0.993 &	0.003 \\
	A2 &	-0.0082 &	0.0025 \\
	x0 &	5.97 &	0.02 \\
	dx &	0.31 &	0.03 \\
\end{tabular}		

Reduced Chi-Sqr	2.64593 $\times$ 10$^{^{-4}}$ \\
Residual Sum of Squares: 0.00159 \\
Adj. R-Square:	0.9989 \\
\vspace{0.5 cm}

L=0.2 m, T = 373 K
\begin{tabular}{ccc}
	Parameter & Value &	Standard Error \\
	A1 &	0.993 &	0.008 \\
	A2 &	-0.0082 &	0.0025 \\
	x0 &	5.97 &	0.02 \\
	dx &	0.31 &	0.03 \\
\end{tabular}		

Reduced Chi-Sqr	2.64587 $\times$ 10$^{^{-4}}$ \\
Residual Sum of Squares: 0.00159 \\
Adj. R-Square:	0.9989	\\
\vspace{0.5 cm}

L=0.2 m, T = 473 K
\begin{tabular}{ccc}
	Parameter & Value &	Standard Error \\
	A1 &	0.993 &	0.008 \\
	A2 &	-0.0082 &	0.0025 \\
	x0 &	5.97 &	0.02 \\
	dx &	0.31 &	0.03 \\
\end{tabular}		

Reduced Chi-Sqr	2.64587 $\times$ 10$^{^{-4}}$ \\
Residual Sum of Squares: 0.00159 \\
Adj. R-Square:	0.9989 \\
\vspace{0.5 cm}

\textbf{A.2}

Errors in the Boltzmann model for the probability calculated according to the truncated normal model.		
\vspace{0.5 cm}
						
L=0.05 m, T = 293 K
\begin{tabular}{ccc}
	Parameter & Value &	Standard Error \\
	A1 &	0.982 &	0.006 \\
	A2 &	-0.0003 &	0.0001 \\
	x0 &	3.29 &	0.01 \\
	dx &	0.24 &	0.01 \\
\end{tabular}		

Reduced Chi-Sqr	6.6611 $\times$ 10$^{^{-5}}$ \\
Residual Sum of Squares: 0.000399 \\
Adj. R-Square:	0.99965 \\
\vspace{0.5 cm}

L=0.05 m, T = 373 K
\begin{tabular}{ccc}
	Parameter & Value &	Standard Error \\
	A1 &	0.982 &	0.006 \\
	A2 &	-0.004 &	0.003 \\
	x0 &	3.19 &	0.02 \\
	dx &	0.22 &	0.02 \\
\end{tabular}	

Reduced Chi-Sqr	6.69947 $\times$ 10$^{^{-5}}$ \\
Residual Sum of Squares: 4.0196 $\times$ 10$^{^{-4}}$ \\
Adj. R-Square:	0.99960 \\
\vspace{0.5 cm}

L=0.05 m, T = 473 K
\begin{tabular}{ccc}
	Parameter & Value &	Standard Error \\
	A1 &	0.982 &	0.005 \\
	A2 &	-0.0004 &	0.0003 \\
	x0 &	3.19 &	0.01 \\
	dx &	0.22 &	0.02 \\
\end{tabular}	

Reduced Chi-Sqr	6.69508 $\times$ 10$^{^{-4}}$ \\
Residual Sum of Squares: 4.01705 $\times$ 10$^{^{-4}}$ \\
Adj. R-Square:	0.99964 \\									
\vspace{0.5 cm}

L=0.1 m, T = 293 K
\begin{tabular}{ccc}
	Parameter & Value &	Standard Error \\
	A1 &	0.991 &	0.003 \\
	A2 &	-0.0022 &	0.0009 \\
	x0 &	5.02 &	0.07 \\
	dx &	0.21 &	0.03 \\
\end{tabular}	

Reduced Chi-Sqr	5.5392 $\times$ 10$^{^{-5}}$ \\
Residual Sum of Squares: 0.00033 \\
Adj. R-Square:	0.99977 \\
\vspace{0.5 cm}

L=0.1 m, T = 373 K
\begin{tabular}{ccc}
	Parameter & Value &	Standard Error \\
	A1 &	0.992 &	0.006 \\
	A2 &	-0.0047 &	0.0025 \\
	x0 &	4.92 &	0.01 \\
	dx &	0.29 &	0.02 \\
\end{tabular}	
 
Reduced Chi-Sqr	1.20178 $\times$ 10$^{^{-4}}$ \\
Residual Sum of Squares: 0.000721 \\
Adj. R-Square:	0.9995 \\
\vspace{0.5 cm}

L=0.1 m, T = 473 K
\begin{tabular}{ccc}
	Parameter & Value &	Standard Error \\
	A1 &	0.995 &	0.005 \\
	A2 &	-0.0045 &	0.0025 \\
	x0 &	4.77 &	0.05 \\
	dx &	0.33 &	0.01 \\
\end{tabular}	

Reduced Chi-Sqr	8.70051 $\times$ 10$^{^{-5}}$ \\
Residual Sum of Squares: 5.2203 $\times$ 10$^{^{-4}}$ \\
Adj. R-Square:	0.99963 \\
\vspace{0.5 cm}

L=0.2 m, T = 293 K
\begin{tabular}{ccc}
	Parameter & Value &	Standard Error \\
	A1 &	0.992 &	0.003 \\
	A2 &	-0.0078 &	0.0065 \\
	x0 &	6.01 &	0.02 \\
	dx &	0.29 &	0.03 \\
\end{tabular}		

Reduced Chi-Sqr	2.75184 $\times$ 10$^{^{-4}}$ \\
Residual Sum of Squares: 0.00165 \\
Adj. R-Square:	0.99885	\\
\vspace{0.5 cm}

L=0.2 m, T = 373 K
\begin{tabular}{ccc}
	Parameter & Value &	Standard Error \\
	A1 &	0.990 &	0.007 \\
	A2 &	-0.0068 &	0.0075 \\
	x0 &	5.99 &	0.02 \\
	dx &	0.28 &	0.03 \\
\end{tabular}		

Reduced Chi-Sqr	2.07471 $\times$ 10$^{^{-4}}$ \\
Residual Sum of Squares: 0.00124 \\
Adj. R-Square:	0.99913	\\
\vspace{0.5 cm}

L=0.2 m, T = 473 K
\begin{tabular}{ccc}
	Parameter & Value &	Standard Error \\
	A1 &	0.993 &	0.008 \\
	A2 &	-0.0082 &	0.0025 \\
	x0 &	5.97 &	0.02 \\
	dx &	0.31 &	0.03 \\
\end{tabular}		

Reduced Chi-Sqr	2.64587 $\times$ 10$^{^{-4}}$ \\
Residual Sum of Squares: 0.00159 \\
Adj. R-Square:	0.9989

\end{document}